\documentstyle[12pt]{article}  
\def\beq{\begin{equation}}  
\def\eeq{\end{equation}}  
\def\bea{\begin{eqnarray}}  
\def\eea{\end{eqnarray}}  
\def\bq{\begin{quote}}  
\def\eq{\end{quote}}  
  
\def\beqa{\begin{eqnarray}}  
\def\eeqa{\end{eqnarray}}  
\def\be{\begin{equation}}  
\def\ee{\end{equation}}  
\def\beq{\begin{equation}}  
\def\eeq{\end{equation}}

\def\bi{\begin{itemize}}  
\def\ei{\end{itemize}}

\newcommand{\barre}[1]{%  
        \setbox1=\hbox{$#1$} \dimen2=\ht1 \dimen3=\dp1 \dimen4=\wd1  
        \setbox2=\hbox{\sl /}  
        \dimen1=\wd1 \advance\dimen1 by -\wd2 \divide\dimen1 by 2  
        \advance\dimen1 by \wd2 \advance\dimen1 by 0.4pt  
        \setbox3=\hbox to \wd1{\hss \box1 \kern -\dimen1 \box2\hss}  
        \ht3=\dimen2 \dp3=\dimen3 \wd3=\dimen4  
        \box3  
        }  
  
\begin{document}  
\pagestyle{empty}  
\begin{flushright}  %  \\  
                     hep-th/0006139 
\end{flushright}  
\vskip 2cm    

\begin{center}  
{\huge The Cosmological Constant Problem  from a Brane-World Perspective\\}  
\vspace*{5mm} \vspace*{1cm}   
\end{center}  
\vspace*{5mm} \noindent  
\vskip 0.5cm  
\centerline{\bf Stefan F\"orste${}^{1}$, Zygmunt Lalak${}^{1,2}$,
St\'ephane Lavignac${}^{1,3}$ and Hans Peter Nilles${}^{1}$
}
\vskip 1cm
\centerline{\em ${}^{1}$Physikalisches Institut, Universit\"at Bonn}
\centerline{\em Nussallee 12, D-53115 Bonn, Germany}
\vskip 0.3cm
\centerline{\em ${}^{2}$Institute of Theoretical Physics}
\centerline{\em Warsaw University, Poland}
\vskip 0.3cm
\centerline{\em ${}^{3}$Service de Physique Th\'eorique, CEA-Saclay}
\centerline{\em F-91191 Gif-sur-Yvette C\'edex, France}
\vskip2cm 
  
\centerline{\bf Abstract}  

We point out several subtleties arising in brane-world scenarios of
cosmological constant cancellation. We show that solutions with curvature
singularities are inconsistent, unless the contribution to the effective
four-dimentional cosmological constant of the physics that resolves the
singularities is fine-tuned. This holds for both flat and curved branes.
Irrespective of this problem, we then study an isolated class of flat
solutions in models where a bulk scalar field with a vanishing potential
couples to a 3-brane. We give an example where the introduction of a bulk
scalar potential results in a nonzero cosmological constant.
Finally we comment on the stability of classical solutions of the brane
system with respect to quantum corrections.

\vskip .3cm

%%%%%%%%%%%%%%%%%%%%%%%%%%%%%%%%%%%%%%%%%%%%%%%%%%%%%%%%%%%%%%%%%%%%%%%%%%%%%  
  
\newpage  

\setcounter{page}{1} \pagestyle{plain}  

\section{Introduction}

With the discovery of string dualities it became clear that extended
objects (branes) are built into any string theory. In the form of
D-branes \cite{Polchinski:1995mt} they first served as tools in
addressing conceptual questions of string theory (e.g.\ dualities,
black hole entropy, holography). An interesting property of D-branes is
that they provide a natural picture for having some fields (typically
gauge fields) being confined to a hypersurface in space whereas others
(typically gravitational fields) can propagate in all directions. This
property makes contact to an early suggestion \cite{Akama:1982jy,
Rubakov:1983bb} that the matter in our universe is confined to live on a 3+1
dimensional hypersurface of a higher dimensional universe.  
In the simplest case the
geometry is such that the higher dimensional space is a direct product
of our 3+1 dimensional universe with some additional internal
space.
This picture was generalized to warped
compactifications \cite{Randall:1999ee}, where the 3+1 dimensional
metric depends on the position of the brane in the higher dimensional
space. The authors of \cite{Randall:1999ee} constructed a model where
this dependence is exponential. They were led to the conjecture that a
natural explanation for the large hierarchy between the electroweak
scale and the Planck scale was found. But even without solving the
hierarchy problem warped compactifications open up interesting
possibilities, like localizing gravity on the brane with an infinitely
large extra dimension \cite{Randall:1999vf}. 

Apart from the hierarchy problem it has been also tried to solve the
cosmological constant problem\footnote{For reviews of the cosmological
constant problem see \cite{Weinberg:1989cp,Witten:2000zk,Binetruy:2000mh}.} 
within the  brane world scenario
\cite{ Arkani-Hamed:2000eg,Kachru:2000hf}. The basic new ingredient
is that there is a scalar living in the bulk with coupling to the
brane. 
Naively one then hopes that the
well-known fine-tuning of the parameters of the theory
required in order
to ensure a vanishing cosmological constant is replaced by the adjustement
of the zero mode of this additional scalar. 
However, these models
typically have singularities being located within a finite
distance from the brane. Problems arising due to the singularities
have for example been pointed out
in \cite{Gubser:2000nd,Forste:2000ps} (this and other aspects of these
models have been also the subject of Refs. \cite{
Chen:2000at,deAlwis:2000qc,deAlwis:2000pr,Csaki:2000wz,Dudas:2000ff,
Horowitz:2000ds,Grinstein:2000fn}). 
In our note \cite{Forste:2000ps} we observed that the singularities can
be given a physical interpretation in the presence of additional source
terms. Then the curvature singularity just reflects the fact that we are
considering the limit in which the  sources (or branes) at the
singularities have no finite
extension into the transverse direction (this is similar to black
hole singularities appearing due to the point like nature of the
source). With these additional sources, however, a fine-tuning is needed
to obtain a vanishing cosmological constant for the effective four
dimensional theory.  Irrespective of this problem, it has been stressed
in \cite{Arkani-Hamed:2000eg,Kachru:2000xs} that for a very
particular choice of parameters  (namely vanishing bulk scalar potential
and particular form and value of the scalar-brane coupling)
there exist isolated flat solutions,
i.e. there seems to be no smooth deformation connecting zero cosmological
constant to a non-vanishing value.

In the present paper we are going to elaborate on  brane world models 
as a framework to find solutions of the cosmological constant problem.
In particular we want to clarify the following points:

\begin{itemize}

\item the vanishing of the cosmological constant requires a consistency
condition for the brane-world setup which is especially crucial
in the presence of singularities (this has been sometimes overlooked
in the literature);

\item known mechanisms to fullfill this consitency condition
require a fine tuning of parameters of the model,
comparable to the usual fine tuning of the cosmological constant.

\end{itemize} 
The so-called self tuning solution of the cosmological constant problem
is therefore at best a scenario where this problem has been
rephrased. The real solution of the problem would have to
explain the fine tuning that is necessary for the consistency condition
mention above.

The paper is organized as follows.
In the next section we specify the general setup and recall consistency
conditions on warped compactifications. In section three we generalize
the points made in \cite{Forste:2000ps}. We show in
various examples (flat and curved branes) that one can make solutions with
singularities consistent by adding additional branes to the model. To
achieve a definite value for the cosmological constant (e.g.\ zero) a
fine-tuning is needed. Section four is devoted to a careful
investigation of the isolated flat solution mentioned above, for 
which we find
that introducing a non trivial bulk potential for the scalar field
necessarily results in a non zero cosmological constant.   
In the fifth section we discuss the relation of the classical solutions
of the brane-bulk system to the physical quantum world.  Finally 
the last section is devoted to a summary and outlook.

%%%%%%%%%%%%%%%%%%%%%%%%%%%%%%%%%%%%%%%%%%%%%%%%%%%%%%%%%%%%%%%%%%%%%%%%%%%%

\section{General Setup}
\label{section:setup}

To study 
thoroughly the structure of vacuum solutions in the brane worlds 
it is useful to consider an ansatz corresponding to maximally symmetric 
4d foliations, which beyond the Minkowski space include de Sitter and 
anti-de Sitter 4d space-times. 
The simplest nontrivial setup   
corresponds to just a single transverse coordinate. The line element is   
\beq \label{ansatz} 
ds^2 = e^{2 A\left( x^5\right)} \tilde{g}_{\mu \nu} dx^\mu dx^\nu +
\left(dx^5 \right)^2 \eeq   
with $ \tilde{g}_{\mu \nu} = \bar{g}_{\mu \nu} + 
h_{\mu \nu}$ where the background metric corresponds to  one of the 
4d maximally symmetric spaces: $\bar{g}_{\mu \nu} = diag \,
\left(-1,+1,+1,+1\right) $
(Minkowski  metric); $diag \, \left(-1,
e^{2 \sqrt{\bar \Lambda} t}, e^{2 \sqrt{\bar \Lambda} t},
e^{2 \sqrt{\bar \Lambda} t}\right)$ ($dS_4$  background with 
$\bar{\Lambda} > 0$); $diag \, \left(- e^{2 \sqrt{- \bar \Lambda} x^3},
e^{2 \sqrt{- \bar \Lambda} x^3},
e^{2 \sqrt{- \bar \Lambda} x^3}, 1\right)$
($AdS_4$ background, $\bar{\Lambda} < 0$).
With $h_{\mu \nu}$ we denote
a small fluctuation around the background metric. 
The sources for such configurations are assumed to be located on    
a number of four dimensional branes, one of which represents the observable   
gauge sector. It is important to note that although the observable gauge   
interactions are strictly confined to the 3-brane, the gravity and moduli   
fields permeate the whole space, effectively connecting the walls in a   
nontrivial way. This implies that every particle localized  on the wall   
feels sources of gravitational forces which are located all over the bulk.   
In some cases the influence of the remote sources will be suppressed, like   
in the case of the exponentially falling off graviton wave function in ref.   
\cite{Randall:1999vf}, which would effectively restrict the relevant   
gravitational sources to the thin layer around the brane, but sometimes   
the suppression would be so mild, that the influence of the whole bulk  
contribution will be highly relevant. Thus in any case, even though the   
gauge forces are restricted to the branes, the gravitational sector has to   
be completely integrated out when going to the effective four dimensional   
theory. In particular, this implies that when one computes the effective   
four dimensional energy density, or four dimensional vacuum pressure, one   
has to integrate over the whole causally accessible portion of the   
transverse space. 

Let us analyze the relation between 4d and 5d physics in some 
detail. 
The 5d action  we consider is (we follow the conventions of Ref.
\cite{Kachru:2000hf} for the normalization of the Einstein term and
of the scalar kinetic term)
\beq \label{eq:genericaction}  
S_5 = \int d^5x \sqrt{-G} \left( R - \frac{4}{3} \left(\partial \phi
  \right)^2  - V\left(\phi\right) \right) 
+ \int_{M_i} d^5 x \sqrt{-g} \left( - f_i \left(\phi\right) \delta \left(x^5 -
  x^5 _i\right)\right)    
\eeq 
where $V (\phi)$ is the bulk scalar potential, $f_i (\phi)$ are brane
contributions to the Lagrangian, the index $i$ counts the
branes, and $g_{\mu \nu}$ is the induced metric on the branes.
The corresponding Einstein equations are  
\beq  \label{eems}
\sqrt{-G} \left( E_{MN} - \frac{1}{2}\, T_{MN} \right) =\ 0\ ,
\eeq
where $E_{MN} = R_{MN} - \frac{1}{2}\, G_{MN} R$ is the Einstein tensor
and
\beq  \label{emt}
T_{MN}\ =\ \frac{8}{3}\, \partial_M \phi \partial_N \phi
- \frac{4}{3}\, ( \partial \phi )^2  G_{MN} - V (\phi) G_{MN}
- f_i \delta \left(x^5\! -\! x^5 _i\right) g_{\mu \nu}
\delta^{\mu}_{M} \delta^{\nu}_{N}
\eeq
is the energy-momentum tensor. The dilaton equation of motion is   
\beq  \label{dem}
- \frac{\partial V}{\partial \phi} \sqrt{-G} - \sqrt{-g}\,  
\delta \left(x^5 - x_i ^5\right) \frac{\partial f_i}{\partial \phi}
+ \frac{8}{3} \partial_M \left( \sqrt{-G} G^{MN} \partial_N \phi
\right) 
= 0,   
\eeq   
where $M,N = 1, \ldots , 5$ and $\mu , \nu = 1, \dots , 4$.  
Although we 
do not consider fluctuations of $G_{55}$
that would correspond in four dimensions to the modulus associated with
the length of the fifth dimension and put here $G_{55}=1$, we shall comment 
on the stability of the fifth dimension in the case of static solutions  
at the end of the paper. With the choice $\bar g_{\mu \nu} = \eta_{\mu \nu}$,
the ansatz (\ref{ansatz}) naively implies that the 4d cosmological constant
should vanish. As explained in \cite{Forste:2000ps} this is the case
if the ansatz 
solves Einstein equations globally, i.e. everywhere in 5d space-time. 
If however there happens to be singularities, for instance at points where 
$e^{A(x^5)}=0$, then one needs to `repair' the system at such points
in order to 
achieve truly flat 4d background. This issue shall be discussed later on.  
In the presence of the general sources the warp factor, hence also the 
curvature scalar, as well as the bulk scalar $\phi$ do assume some nontrivial 
and $y$-dependent vacuum configuration.

To identify 4d gravity and the strenght of its coupling to matter one 
needs to split the action (\ref{eq:genericaction}) into vacuum part, 
and fluctuations around 
it. The vacuum part is read off from the 5d action upon substituting 
solutions of the equations of motion for the warp factor and for the scalar. 
This vacuum Lagrangian is  
\beq
\left< L_5\right> = \int dy\, e^{4 \left<A(x^5)\right> } \left( \frac{2}{3} 
V\left(\left<\phi \right>\right) 
+ \frac{1}{3} \left< f_i \right> \delta \left(x^5 - x^{5}_i\right)
\right) , 
\eeq
where $\left< \ldots\right>$ indicates that the classical background
value of the corresponding field is taken.
It is immediate to see that the lowest terms in $h_{\mu \nu} $ of the
expansion of the action (\ref{eq:genericaction}) around the vacuum are 
\beq \label{mpl}
S_5 = \int d^5 x \sqrt{-\tilde{g}}  
M^3 e^{2 A(x^5)} \tilde{R} + ...
\eeq
where $\tilde{R}$ denotes the curvature built out of $\tilde{g}_{\mu \nu} $.
From (\ref{mpl}) one can identify the 4d Planck scale $M^{2}_{Pl} =
M^3 \int dy e^{ 2 <A(x^5)>}$. To obtain the effective four
dimensional action $S_4$ for gravity one needs to integrate over the fifth
coordinate. The model is consistent if
$\left<\tilde{g}_{\mu\nu}\right>$ from our ansatz (\ref{ansatz})
minimizes $S_4$. Thus the effective four dimensional action reads
\begin{equation}
S_4 = M_{Pl}^2\int d^4x\sqrt{-\tilde{g}}\left( \tilde{R} - \lambda\right)\ , 
\end{equation}
where the cosmological constant $\lambda = 6 \bar \Lambda$.
Finally we arrive at a consistency condition by the requirement that the
on-shell values of $S_5$ and $S_4$ should be equal\footnote{Note that
for $\bar{\Lambda}=0$ this is just the condition of vanishing
vacuum energy imposed in \cite{Forste:2000ps}.}
\begin{equation} \label{eq:condition}
\left< L_5 \right>\ =\ M^2_{Pl} \left( \langle \tilde{R} \rangle - \lambda
\right)\ =\ 6 \bar{\Lambda} M_{Pl}^2\ .
\end{equation}
A breakdown of this condition would signal that Einstein equations are not
satisfied everywhere in space-time. It is interesting to note
that Eq. (\ref{eq:condition}) can be rewritten
\begin{equation}
  -\, \frac{1}{3} \int dx^5\, e^{4 <A>}\,
  \langle T^0_0 + T^5_5 \rangle\ =\  6 \bar \Lambda M^2_{Pl}\ .
\label{eq:lambda_T_MN}
\end{equation}
In the case of a Poincar\'e-invariant background, 
the vanishing of the 4d cosmological constant is therefore equivalent to a
constraint on the 5d energy-momentum tensor, namely the integral on the
left-hand side of (\ref{eq:lambda_T_MN}) should vanish. This constraint is a
variant of the condition
$\int dx^5 e^{<A>} \langle T^0_0 - \frac{1}{2} T^5_5 \rangle = 0$
derived in Ref. \cite{Ellwanger:2000pq}, and it has the same origin: the
combination $e^{4<A>} \langle T^0_0 + T^5_5 \rangle$ is constrained by Einstein
equations to be a total derivative, therefore its integral over a compact
interval should vanish. However, while
$\int dx^5 e^{<A>} \langle T^0_0 - \frac{1}{2} T^5_5 \rangle = 0$ also holds
for a $dS_4$ or $AdS_4$ background, the condition
$\int dx^5 e^{4<A>} \langle T^0_0 + T^5_5 \rangle = 0$ is more specific for
Poincar\'e invariance. To see how Eq. (\ref{eq:lambda_T_MN}) (with
$\bar \Lambda = 0$) relates the vanishing of the 4d cosmological constant
to the consistency of the 5d Einstein equations, let us note
that in the Randall-Sundrum model \cite{Randall:1999ee}
$r_c \int dx^5 e^{4<A>} \langle T^0_0 + T^5_5 \rangle =
(e^{-4 \pi k r_c} - 1)\, \Lambda/k - e^{-4 \pi k r_c} V_{obs} - V_{hid}$\ ;
therefore, the above constraint is 
automatically satisfied once the fine-tuning needed to achieve a
Poincar\'e-invariant solution in four dimensions (namely
$V_{hid} = - V_{obs} = \Lambda/k$) is imposed. A similar cancellation is
expected to appear in the approach put forward in Ref.
\cite{Rubakov:1983bb,Verlinde:2000xm,Schmidhuber:1999rc}.

%%%%%%%%%%%%%%%%%%%%%%%%%%%%%%%%%%%%%%%%%%%%%%%%%%%%%%%%%%%%%%%%%%%%%%%%%%%%%

\section{Vacuum configurations with singularities and fine-tuning}

In this section, we consider various vacuum configurations leading to
singular warped compactifications, and show that in this case Einstein
equations are not globally satisfied (condition (\ref{eq:lambda_T_MN})
is not fulfilled), leading to an inconsistency of the solution.
This implies in particular that the
recently proposed ``self-tuning'' mechanism of the cosmological constant 
\cite{Arkani-Hamed:2000eg,Kachru:2000hf} does not work as it stands,
and hides a fine-tuning at the singularity.

Let us first recall what is generally understood under ``self-tuning of the
cosmological constant'' in brane-world models.
It is the property that solutions of the five-dimensional Einstein equations
preserving Poincar\'e invariance on the brane can be found for a wide
range of values of the brane tension, i.e. independently of the brane
contribution to the effective cosmological constant. 
This behaviour, first identified in Ref. \cite{Arkani-Hamed:2000eg} and 
\cite{Kachru:2000hf}, has been shown to be generic in models where a single
brane is coupled to gravity and a scalar field \cite{Csaki:2000wz}.
Indeed, plugging the ansatz
\begin{equation}
  ds^2\ =\ e^{2 A(x^5)} \eta_{\mu\nu}\, dx^\mu dx^\nu\ +\ 
  \left( dx^5 \right)^2
\label{eq:Poincare_ansatz}
\end{equation}
into the equations of motion (\ref{eems}) and (\ref{dem}) (which are now
written for a single brane located at $x^5=0$), one generally finds
solutions without imposing any relation among the parameters of the
Lagrangian. In particular, the brane tension $f (\phi)$ - which contains the
contribution of the fields living on the brane (the SM fields) to the
four-dimensional cosmological constant - needs not be fine-tuned and may vary
over a wide range of values.

However, these solutions either do not localize gravity on the
brane (the graviton zero mode is not a normalizable bound state, or
equivalently the four-dimensional Planck mass diverges) or they have curvature
singularities at finite proper distance\footnote{The authors of Ref.
\cite{Csaki:2000wz} have shown that in this class of models, the only
solutions that localize gravity with an infinitely large extra dimension
involve a fine-tuning between bulk and brane parameters, like in the
Randall-Sundrum model.}. At these singularities, the warp factor vanishes
- implying that the four-dimensional metric degenerates, 
$G_{\mu \nu} \rightarrow 0$ - and the curvature scalar diverges,
signaling a breakdown of Einstein gravity. In the absence of any mechanism
that would smooth these singularities, one assumes that their effect is
simply to cut off the fifth dimension, effectively compactifying it to a
finite interval and ensuring localization of gravity on the brane. 
However, due to the presence of singularities the consistency condition
(\ref{eq:lambda_T_MN}) is not satisfied, which indicates that the ansatz
(\ref{eq:Poincare_ansatz}) does not solve the equations of motion
on the whole interval. This can be seen in explicit examples by the
fact that the effective four-dimensional cosmological constant,
Eq. (\ref{eq:condition}),
\begin{equation}
  \lambda\, M^2_{Pl}\ =\ -\, \frac{2}{3} \int dx^5\, e^{4A}\, V (\langle \phi
  \rangle)\ -\ \frac{1}{3}\: e^{4A} f (\langle \phi \rangle)
  \left| _{x^5=0} \right.
\label{eq:lambda_Poincare}
\end{equation}
does not vanish, in spite of the Poincar\'e invariance of the geometry on the
brane. Thus, in order for the solution to make sense, the (unknown)
microphysics that resolves the singularity has to
contribute to $\lambda$ in such a way that it exactly cancels the
brane and bulk contributions. This non-trivial statement ruins the
apparent self-tuning behaviour of the solution: the traditional fine-tuning
problem of the cosmological constant has simply been shifted into an unknown
sector of the model. Of course one cannot exclude that an adjustement
mechanism, possibly involving some new light degrees of freedom
\cite{Kachru:2000xs,Csaki:2000wz}, may restore the self-tuning; but a concrete
realization of this idea is still missing.

In order to make the problem explicit, one can parametrize the contribution
of the unknown physics at the singularity by adding delta source terms
in the energy-momentum tensor, $T_{MN} = - \sum_s f_s (\phi)
\delta (x^5 - x_s) g_{\mu \nu} \delta^{\mu}_M \delta^{\nu}_N + \ldots$,
where $x_s$ denotes the position of a singularity.
The necessity of adjusting precisely these terms for each particular
solution (for each value of the brane tension) spoils the self-tuning. Indeed,
in addition to the boundary conditions at the brane, the solution has to
satisfy, at each singularity $x_s$, the following two constraints:
\begin{eqnarray}
  e^{4A(x_s)} \left[ \phi' (x_s+0) - \phi' (x_s-0) \right] & = &
  \frac{3}{8}\ e^{4A(x_s)}\, \frac{\partial f_s}{\partial \phi}
  [\phi(x_s)]\ ,
  \label{eq:BC_singularity_1}  \\
  e^{6A(x_s)} \left[ A' (x_s+0) - A' (x_s-0) \right] & = &
  -\, \frac{1}{6}\ e^{6A(x_s)} f_s[\phi(x_s)]\ .
  \label{eq:BC_singularity_2}
\end{eqnarray}
The fine-tuning implied by Eq. (\ref{eq:BC_singularity_1}) and
(\ref{eq:BC_singularity_2}) guarantees that the consistency condition
(\ref{eq:lambda_T_MN}) is fulfilled and, as a consequence, that the 4d
cosmological constant vanishes.

In Ref. \cite{Forste:2000ps}, this point was illustrated in the case of a
vanishing bulk scalar potential, $V (\phi) = 0$, addressed in Ref.
\cite{Arkani-Hamed:2000eg} and \cite{Kachru:2000hf}. In this section, we
would like to discuss another example and to show that the requirement of
having a global solution of the five-dimensional equations of motion in the
presence of curvature singularities imposes a fine-tuning for more general
ans\"atze than the (four-dimensional) Poincar\'e-invariant background
(\ref{eq:Poincare_ansatz}). 

Let us first consider the case of a flat brane located at
$x^5=0$, i.e. we take the Poincar\'e-invariant ansatz
(\ref{eq:Poincare_ansatz}). In order to illustrate our point, we choose
examples of bulk scalar potentials for which one can obtain a
simple analytical expression for the solution of the equations of motion.
The simplest case corresponds to a vanishing bulk potential and has been
addressed in Ref. \cite{Forste:2000ps}. Let us repeat shortly the discussion
here. The solution of the bulk equations reads \cite{Kachru:2000hf}
\begin{equation}  
 \phi \left( x^5 \right) = \left\{  
 \begin{array}{ll}  
  \frac{3}{4}\, \epsilon_1 \log \left| \frac{4}{3} x^5 + c_1 \right| + d_1\ ,
  & x^5 < 0   \\  
  \frac{3}{4}\, \epsilon_2 \log \left| \frac{4}{3} x^5 + c_2 \right| + d_2\ ,
  & x^5 >0 
 \end{array} \right.\ ,
\label{eq:Phi_zero}
\end{equation}
\begin{equation}
 A' \left( x^5 \right) = \left\{  
 \begin{array}{ll}
  \frac{1}{3}\, \epsilon_1\, \phi' \left( x^5 \right)\ ,  &  x^5 < 0  \\
  \frac{1}{3}\, \epsilon_2\, \phi' \left( x^5 \right)\ ,  &  x^5 > 0
 \end{array} \right. ,
\label{eq:A_zero}
\end{equation}
where $\epsilon_{1,2} = \pm 1$ and $c_1$, $c_2$, $d_1$ and $d_2$ are
integration constants. The boundary conditions on the brane determine
three of them in terms of the fourth one and of the parameters in the
Lagrangian; this means in particular that there are solutions for a wide
range of values of the brane tension and of the scalar coupling to the brane.
The $x^5$-dependence of the warp factor
($e^{2A(x^5)} \propto |\frac{4}{3} x^5 + c_i|^{1/2}$) does not allow for a
localization of gravity on the brane with an infinite extra dimension;
therefore, the only phenomenologically acceptable solutions are the ones that
have singularities on both axes $x^5<0$ and $x^5>0$, effectively compactifying
the fifth dimension to a finite interval. In practice this means that we
must choose $c_1>0$ and $c_2<0$; the singularities are then located at
$x_- = - \frac{3}{4} c_1$ and $x_+ = - \frac{3}{4} c_2$. Following Ref.
\cite{Forste:2000ps}, we truncate the solution at the
singularities, i.e. we set $|x^5 - x_-| = 0$ for $x^5 < x_-$ and
$|x^5 - x_+| = 0$ for $x^5 > x+$. As explained above, (\ref{eq:Phi_zero})
and (\ref{eq:A_zero}) do not solve the equations of motion on the closed
interval $[x_-,x_+]$. This is reflected in the fact that, as can be
immediately seen from Eq. (\ref{eq:lambda_Poincare}), the effective
four-dimensional cosmological constant does not vanish, unless the brane
tension itself vanishes. A simple way to cure this inconsistency is to
``resolve'' the singularities by adding the following source terms to the
action (\ref{eq:genericaction}):
\begin{equation}
  - \int d^4x\, \sqrt{-g}\, f_- (\phi) \left|_{x^5=x_-} \right.\
  - \int d^4x\, \sqrt{-g}\, f_+ (\phi) \left|_{x^5=x_+} \right.\ .
\end{equation}
The matching conditions (\ref{eq:BC_singularity_1}) and
(\ref{eq:BC_singularity_2}) then amounts to a fine-tuning of the source terms
at the singularities. Assuming for example exponential couplings for $\phi$
as in \cite{Arkani-Hamed:2000eg,Kachru:2000hf}, $f(\phi) = e^{b \phi}\, T$,
$f_-(\phi) = e^{b_- \phi}\, T_-$ and $f_+(\phi) = e^{b_+ \phi}\, T_+$,
one obtains
\begin{equation}
 \begin{array}{lll}
  b_- = - \frac{4}{3}\ ,  &  &  T_- = - 2\, e^{+ \frac{4}{3} d_1}\ ,  \\
  b_+ = + \frac{4}{3}\ ,  &  &  T_+ = - 2\, e^{- \frac{4}{3} d_2}\ ,
 \end{array}
\label{eq:f_s_zero_I}
\end{equation}
and $\epsilon_1 = - \epsilon_2 = +1$ for $|b| < \frac{4}{3}$ (solution (I)
of Ref. \cite{Kachru:2000hf}), and
\begin{equation}
 \begin{array}{lll}
  b_- = b_+ = - \frac{4}{3}\ ,  &  &  T_- = T_+ = - \frac{T}{2}\ ,
 \end{array}
\label{eq:f_s_zero_II}
\end{equation}
and $\epsilon_1 = \epsilon_2 = +1$ for $|b| = \frac{4}{3}$ (case considered
in Ref. \cite{Arkani-Hamed:2000eg} and solution (II)
of Ref. \cite{Kachru:2000hf}). Note that since in the case $|b| = \frac{4}{3}$
the solution is symmetric under $x^5 \leftrightarrow - x^5$, it is possible to
identify the two singularities and to treat $x^5$ as a periodic coordinate
(this amounts to continue periodically the solution beyond the singularities
instead of ``truncating'' it as we did above). Then one needs to add a single
energy source at $x_s \equiv x_- = x_+$, $- \int d^4x\, \sqrt{-g}\,
e^{b_s \phi} T_s \left|_{x^5=x_s} \right.$  with $b_s = - \frac{4}{3}$ and
$T_s = - T$.

The second explicit example \cite{Kanti:2000rd} we would like to discuss
corresponds to a constant bulk scalar potential, i.e. a nonvanishing bulk
cosmological constant, $V(\phi) = \Lambda_B$. For the sake of simplicity,
we focus on the case $\Lambda_B < 0$ and consider only $Z_2$-symmetric
solutions; when convenient we shall work on the half line $x^5 \geq 0$.
The solution of the bulk equations reads \cite{Kanti:2000rd}
\begin{equation}  
  e^{4A(x^5)}\ =\ e^{4A(0)} \left[\, \cosh (\omega |x^5|)
  + r \sinh (\omega |x^5|)\, \right]\ ,
\end{equation}
\begin{equation}
  \phi'(x^5)\ =\ c\, e^{-4 [A(x^5)-A(0)]} \qquad
  \phi(x^5)\ =\ c \int^{x^5}_{0} \! \! \! dy\,
  e^{-4 [A(y)-A(0)]}\, +\, d\ ,
\end{equation}
\begin{equation}  
  c^2\ =\ \frac{9}{16}\ \omega^2 \left[\, r^2
  - 1\, \right]\ ,
\label{eq:eom_Lambda_B_3}
\end{equation}
where $\omega \equiv \sqrt{- \frac{4}{3}\, \Lambda_B}$. Note that the last
equation requires $|r| \geq 1$. The boundary conditions
on the brane give the following two relations
\begin{equation}
  f[\phi(0)]\ =\ -3\, \omega\, r\ ,  \qquad \qquad
  \frac{\partial f}{\partial \phi}\, [\phi(0)]\ =\ \frac{16}{3}\: c\ .
\label{eq:boundary_Lambda_B}
\end{equation}
Plugging (\ref{eq:boundary_Lambda_B}) into (\ref{eq:eom_Lambda_B_3}), one
obtains a relation between $f[\phi(0)]$, $\frac{\partial f}{\partial \phi}\,
[\phi(0)]$ and $\Lambda_B$. Given $f(\phi)$ and $\Lambda_B$, it is
generally possible to find a value of $d \equiv \phi(0)$ which satisfies this
constraint for a wide range of values of the brane tension $f[\phi(0)]$; Eq.
(\ref{eq:boundary_Lambda_B}) then determines the other two integration
constants $r$ and $c$. For instance, if $\phi$ has only exponential
couplings, $f(\phi) = e^{b \phi}\, T$,
$\phi(0)$ is determined to be $\phi(0) = \frac{1}{2b}\,
\ln \left[16 \omega^2 / T^2 (\frac{16}{9} - b^2) \right]$ (note the
restriction on the parameter $b$, $|b| < \frac{4}{3}$).
Therefore, this solution has a self-tuning behaviour;
however, it localizes gravity on
the brane only if $r \leq - 1$. In the case $r < - 1$,
there is a singularity at finite proper distance, located at
\begin{equation}
  x_s\ =\ \frac{1}{\omega}\: \mbox{Arctanh} \left( - \frac{1}{r} \right)
\label{eq:singularity_Lambda_B}
\end{equation}
(together with its twin singularity at $x^5 = -x_s$), and the solution can be
rewritten
\begin{equation}  
  e^{4A(x^5)}\ =\ e^{4A(0)}\: \frac{\sinh \left[\omega (x_s - |x^5|) \right]}
  {\sinh (\omega x_s)}\ ,
\end{equation}
\begin{equation}
  \phi(x^5)\ =\ -\, \frac{3}{4}\, \epsilon\, \log \left(\,
  \frac{\tanh \left[\, \frac{\omega}{2} (x_s - |x^5|)\, \right]}
  {\tanh \left( \frac{\omega x_s}{2} \right)} \right)\,  +\, d\ ,
\end{equation}
where $\epsilon = \mbox{sign} \left( \frac{\partial f}{\partial \phi}
[\phi(0)] \right)$. In the case $r = - 1$, the solution becomes
$e^{4A(x^5)} = e^{4A(0)} e^{-\omega |x^5|}$ and the singularity is pushed to
infinity. This limit corresponds to a constant $\phi$ background
(i.e. $\phi(x^5) = \mbox{cst}$, as implied by Eq. (\ref{eq:eom_Lambda_B_3})),
and Eq. (\ref{eq:boundary_Lambda_B}) amounts to a fine-tuning between the bulk
cosmological constant $\Lambda_B$ and the brane tension
$V_0 \equiv f[\phi(0)]$, namely $V_0 = 3\, \omega$. It is interesting to
note \cite{Kanti:2000rd} that this fine-tuning is precisely the one appearing
in the Randall-Sundrum model (here the configuration would be the one
considered in Ref. \cite{Randall:1999vf}, in which the second brane is pushed
to infinity; note that our notations correspond to setting $2 M = 1$ in the
formula of Randall and Sundrum). This difference of behaviour of the two
solutions (fine-tuning versus self-tuning) is already striking, given the fact
that the fine-tuned solution can be obtained from the ``self-tuning'' solution
by taking the limit $x_s \rightarrow \infty$. If one now computes the
effective four-dimensional cosmological constant, one finds, using
(\ref{eq:lambda_Poincare}),
\begin{equation}
  \lambda\, M^2_{Pl}\ =\ -\, \frac{e^{4A(0)}\, \omega}{\sinh (\omega x_s)}\ .
\label{eq:lambda_Lambda_B_incomplete}
\end{equation}
Thus, in the case $r < -1$ (``self-tuning'' solution with a singularity at
finite proper distance), there is only a partial cancellation between the
bulk and the brane contributions to the 4d cosmological constant, while in
the case $r = -1$ (fine-tuned solution with a singularity at infinity)
$\sinh (\omega x_s) = \infty$ and therefore $\lambda = 0$. This resolves the
apparent paradox noted above: the ``self-tuning'' solution is incomplete, i.e.
an additional source has to be added at the singularity in order for the
equations of motion to be globally satisfied. Specifically, adding
$- \int d^4x\, \sqrt{-g}\, e^{b_s \phi} T_s \left|_{x^5=x_s} \right.$ to the
action, one finds that a fine-tuning of $b_s$ and $T_s$ is required,
\begin{equation}
  b_s\ =\ \frac{4}{3}\: \epsilon\ ,  \qquad
  T_s\ =\ -\, \frac{3}{2}\: \frac{\omega\, e^{-b_s d}}
  {\tanh \left(\frac{\omega x_s}{2}\right)}\ .
\label{eq:f_s_Lambda_B}
\end{equation}
The contribution of the singularity to the 4d cosmological constant
then exactly cancels (\ref{eq:lambda_Lambda_B_incomplete}), as expected. In
the limit $x_s \rightarrow \infty$, this contribution vanishes.

Another way to understand the necessity of adding an energy source at the
singularity is to consider a slightly different set-up, in which a second
``end of the world'' brane is placed at a position $x^5 = L < x_s$. Then Eq.
(\ref{eq:boundary_Lambda_B}) must be supplemented with a set of
new boundary conditions at $x^5 = L$,
\begin{eqnarray}
  f_L[\phi(L)] & = & -3\, \omega\, \coth \left[\omega (x_s - L)\right]\ ,
\label{eq:boundary_L_Lambda_B_1}  \\
  \frac{\partial f_L}{\partial \phi}\, [\phi(L)] & = &
  -\, \frac{16}{3}\ c\, \frac{\sinh (\omega x_s)}
  {\sinh \left[ \omega (x_s - L) \right]}\ .
\label{eq:boundary_L_Lambda_B_2}
\end{eqnarray}
Since the integration constants $d$, $r$ (hence $x_s$) and $c$ are
determined
by the boundary conditions at $x^5 = 0$, Eq. (\ref{eq:boundary_L_Lambda_B_1})
and (\ref{eq:boundary_L_Lambda_B_2}) imply a fine-tuning of $f_L (\phi)$,
which at the same time fixes the inter-brane distance $L$, in the
same way as in Ref. \cite{Goldberger:1999uk}. More specifically, for an
exponential coupling of the bulk scalar field $f_L(\phi) = e^{b_L \phi}\, T_L$,
Eq. (\ref{eq:boundary_L_Lambda_B_1}) and (\ref{eq:boundary_L_Lambda_B_2})
yield
\begin{equation}
  b_L\ =\ \frac{4}{3}\: \epsilon\, \cosh^{-1} \left[\omega (x_s\! -\! L)
  \right]\ ,
\label{eq:b_L_Lambda_B}
\end{equation}
\begin{equation}
  T_L\ =\ -3\, \omega\, e^{-b_L d} \coth \left[\omega (x_s\! -\! L)\right]
  \left( \frac{\tanh \left[\, \frac{\omega}{2} (x_s\! -\! L)\, \right]}
  {\tanh \left( \frac{\omega x_s}{2} \right)}
  \right)^{\cosh^{-1} \left[\omega (x_s\! -\! L)\right]}\ .
\label{eq:T_L_Lambda_B}
\end{equation}
Thus, for a given value of $b_L$ (resp. $T_L$), Eq. (\ref{eq:b_L_Lambda_B})
(resp. Eq. (\ref{eq:T_L_Lambda_B})) determines the inter-brane distance $L$,
while Eq. (\ref{eq:T_L_Lambda_B}) (resp. Eq. (\ref{eq:b_L_Lambda_B})) tells us
that the brane vacuum energy $T_L$ (resp. the scalar coupling $b_L$) must be
fine-tuned. Conversely, for a fixed value of $L$, both $b_L$ and $T_L$ must be
fine-tuned. If one now sends the second brane to the singularity,
$L \rightarrow x_s$, one finds exactly $b_L \rightarrow b_s$ and
$T_L \rightarrow T_s$: the ``self-tuning'' single-brane set-up is
nothing but the (singular) limit of a fine-tuned two-brane system, in which the
second brane hides at the singularity\footnote{It is interesting to consider
another limit, in which the singularity is pushed to infinity, while the
second brane remains at a finite proper distance $L$ from the brane at the
origin. Then one recovers again the Randall-Sundrum model (in the configuration
of Ref. \cite{Randall:1999ee}), with the double fine-tuning
$V_0 = - V_L = 3\, \omega$, where $V_L \equiv \lim_{x_s \rightarrow \infty}
f_L [\phi(L)]$.}.

Note that we could ``screen'' the singularities in the same way in the case of
a vanishing scalar potential discussed previously, i.e. we could place two
branes at $x^5= L_- > x_-$ and at $x^5= L_+ < x_+$. Unlike what we have just
found for a constant scalar potential, we would then conclude that the brane
tensions and scalar couplings do not depend on the positions of the two
additional branes, i.e. Eq. (\ref{eq:f_s_zero_I}) (resp. Eq.
(\ref{eq:f_s_zero_II}) in the $Z_2$ symmetric case) hold for any inter-brane
distances, and $L_-$, $L_+$ are moduli. The statement that one recovers the
``self-tuning'' solution in the limit $L_- \rightarrow x_-$,
$L_+ \rightarrow x_+$ follows then trivially.

We could go on discussing other examples with a more general bulk scalar 
potential; however, while the explicit form of the solutions would become
more involved, the conclusions would remain the same.

We would now like to show that the necessity of adding energy sources
in order to ensure the consistency of a solution of Einstein equations
with singularities is not a particularity of the ansatz
(\ref{eq:Poincare_ansatz}). For this purpose, we come back to the case of
a vanishing bulk scalar potential, in which curved solutions with maximal
symmetry (de Sitter or anti-de Sitter) in four dimensions can be found
\cite{Kachru:2000xs}. Our ansatz is
\begin{equation}
  ds^2\ =\ e^{2 A(x^5)}\, \bar g_{\mu\nu}\, dx^\mu dx^\nu\ +\ 
  \left( dx^5 \right)^2
\label{eq:curved_ansatz}
\end{equation}
where $\bar g_{\mu \nu}$ is the metric of a maximally symmetric $3+1$
dimensional space with a curvature constant $\bar \Lambda$
(the explicit form of $\bar g_{\mu \nu}$ has been given at the beginning of
Section \ref{section:setup}). With this ansatz, the solution of the bulk
equations of motion can be
written in terms of a hypergeometric function $F(z) \equiv \ _{2} \! F_{1}
(\frac{1}{2}, \frac{2}{3}, \frac{5}{3}, z)$ \cite{Kachru:2000xs}
\begin{equation}
  \frac{1}{\gamma_i}\: e^{4A(x^5)}\, F \left(-\frac{9 \bar \Lambda}
  {\gamma_i^2}\: e^{6A(x^5)} \right)\ =\
  \epsilon_i \left( \frac{4}{3}\, x^5 + c_i \right)\ ,
\end{equation}
\begin{equation}
  \phi'(x^5)\ =\ \gamma_i\, e^{-4A(x^5)}\ ,
\end{equation}
where $i=1,2$ refer to $x^5<0$ and $x^5>0$ respectively,
$\epsilon_{1,2} = \pm 1$, $c_1$, $c_2$ are integration constants and
$\gamma_i = \epsilon_i F|_{x^5=0}/ c_i$ if we choose $A(0)=0$. The boundary
conditions on the brane give the following two constraints,
\begin{equation}
  f[\phi(0)]\ =\ 2 \left( \sqrt{\gamma^2_1 + 9 \bar \Lambda} +
  \sqrt{\gamma^2_2 + 9 \bar \Lambda} \right)\ ,  \qquad
  \frac{\partial f}{\partial \phi}\, [\phi(0)]\ =\ \frac{8}{3}\:
  (\gamma_2 - \gamma_1)\ .
\label{eq:boundary_0_curved}
\end{equation}
Thus, given $f(\phi)$, one generally finds a continuous set of
solutions corresponding to different values of $\bar \Lambda$ (including
a flat solution $\bar \Lambda = 0$). The authors of Ref.
\cite{Arkani-Hamed:2000eg} and \cite{Kachru:2000xs} have shown that
for an exponential coupling of the scalar field to the brane, there is a
particular value of the coupling which does not allow for any curved solution.
Indeed, pluging $f(\phi) = e^{b \phi}\, T$ in the above constraints
(\ref{eq:boundary_0_curved}), one sees that only $\bar \Lambda = 0$ leads
to a solution in the case $b = \pm \frac{4}{3}$, while any $\bar \Lambda$ is
allowed for $|b| < \frac{4}{3}$ (and any $\bar \Lambda < 0$ is allowed for
$|b| > \frac{4}{3}$). Here we are interested in curved solutions, so we choose
$b \neq \pm \frac{4}{3}$ and consider $\bar \Lambda \neq 0$. The solutions
have singularities on both axes if we take $c_1>0$ and $c_2<0$, like in the
$\bar \Lambda = 0$ case. The vacuum energy is given by Eq.
(\ref{eq:lambda_Poincare}) with an additional term $12 \bar \Lambda M^2_{Pl}$
on the RHS; assuming first that the singularities do not contribute, one finds
\begin{equation}
  \lambda M^2_{Pl}\ =\ - \frac{2}{3}\, \left(\sqrt{\gamma^2_1 + 9 \bar \Lambda}
  + \sqrt{\gamma^2_2 + 9 \bar \Lambda}\right)\ +\ 12 \bar \Lambda M^2_{Pl}\ ,
\end{equation}
while the Planck mass is computed to be
\begin{equation}
  M^2_{Pl}\ =\ \frac{1}{9 \bar \Lambda} \left[ \left(\sqrt{\gamma^2_1
  + 9 \bar \Lambda} + \sqrt{\gamma^2_2 + 9 \bar \Lambda}\right)
  - \left(|\gamma_1| + |\gamma_2|\right) \right]\ .
\end{equation}
Since $\bar R = 12 \bar \Lambda$, the four-dimensional Einstein equations
require $\lambda = 6 \bar \Lambda$. Clearly this is not the case here, which
indicates an inconsistency of the
solution. Adding now delta energy sources $f_{\pm}(\phi)$ at the
singularities, one obtains two new sets of two boundary conditions,
\begin{eqnarray}
  \left. e^{4A} f_{\pm}(\phi)\, \right|_{x^5=x_{\pm}} & = &
  - 2\, |\gamma_{|^2_1}|\ ,  
\label{eq:boundary_s_curved}  \\
  \left. e^{4A} \frac{\partial f_{\pm}}{\partial \phi} (\phi)\,
  \right|_{x^5=x_{\pm}} & = & \mp\, \frac{8}{3}\: \gamma_{|^2_1}\ ,
\end{eqnarray}
where $\gamma_{|^2_1} = \gamma_1$ for $x^5=x_-$ and $\gamma_2$ for $x^5=x_+$.
It is now straightforward, using Eq. (\ref{eq:boundary_s_curved}), to check
that the singularities give a contribution $\frac{2}{3} \left(|\gamma_1|
+ |\gamma_2|\right)$ to $\lambda M^2_{Pl}$, therefore ensuring
$\lambda = 6 \bar \Lambda$, as required by the 4d
Einstein equations. Note that, like in the $\bar \Lambda = 0$ case, the
boundary conditions at the singularities introduce a fine-tuning which
is naively absent if one does not insist on the validity of Einstein
equations at the singularities. Thus even curved
solutions require a fine-tuning.

%%%%%%%%%%%%%%%%%%%%%%%%%%%%%%%%%%%%%%%%%%%%%%%%%%%%%%%%%%%%%%%%%%%%%%%%%%%%%%

\section{More on nearby curved solutions}

In the previous section we have seen that solutions whith
singularities are not consistent unless one specifies the physics at
the singularity. Our proposal was to interprete the corresponding
curvature singularities as arising due to the presence of additional
source terms. This proposal passed the consistency condition
(\ref{eq:condition}). The lesson we learned is that fine tuning
on the parameters of sources at the singularities is
needed. Therefore, these setups do not solve the cosmological constant
problem. However, irrespective of the fine-tuning problem, they possess
an additional property which makes them attractive in view of the cosmological
constant problem: for the very specific choice of parameters (namely vanishing
bulk potential and particular value of the scalar-brane coupling) discussed
in \cite{Arkani-Hamed:2000eg} and in solution (II) of \cite{Kachru:2000hf}, 
there are no nearby curved solutions \cite{Arkani-Hamed:2000eg,Kachru:2000xs}.
In the present section we are going to rederive that result in a simpler way,
and afterwards we will explore what happens when one adds a bulk potential
for the scalar field $\phi$.

The action considered in \cite{Arkani-Hamed:2000eg,Kachru:2000hf} reads
\begin{equation}
S= \int d^5x\sqrt{-G}\left[ R - \frac{4}{3}\left(\nabla \phi\right)^2\right]
-\int_{x^5=0}d^4x \sqrt{-g}T e^{{b}\phi} .
\end{equation}
In our previous notations (\ref{eq:genericaction}), this corresponds to
$V(\phi) = 0$ and $f(\phi) = T e^{b \phi}$.
This is some version of five dimensional Jordan-Brans-Dicke (JBD) theory in
the Einstein frame where the scalar $\phi$ couples in a universal
manner (as a conformal factor of the metric) to matter. However, all
matter is confined to live on a 4d hypersurface located at $x^5
=0$. The metric on this hypersurface is taken as the induced one.
We define the JBD frame by performing a Weyl transformation such that
$\phi$ decouples from all matter. 
This is achieved by the field redefinition
$G_{MN}\rightarrow e^{-b/2 \phi}G_{MN}$. In the JBD frame the action reads,
\begin{equation} \label{brane-frame}
S= \int d^5 x  \sqrt{-G}e^{-\frac{3}{4}b\phi}\left[ R +\left(\frac{3}{4}b^2
    -\frac{4}{3}\right) \left(\nabla\phi\right)^2\right]
-\int_{x^5=0}d^4x\sqrt{-g} T .
\end{equation}
Now, for $b=\pm \frac{4}{3}$ we observe that the $\phi$ equation of
motion results in the constraint
$R=0$
and we see that this case is special. 
It corresponds exactly to the setup where only flat solutions
exist \cite{Arkani-Hamed:2000eg,Kachru:2000xs}. 
Now, we will re-derive this result in the JBD frame (\ref{brane-frame})
where the calculation turns out to simplify substantially. This will allow us
to study how the solution is affected by the presence of a bulk scalar
potential.
We should remark that in this section we understand that, in the
spirit of the previous section, sources at 
singularities need to be added. However, in order to keep formulas
short we do not write down those terms explicitly.
Typically we are looking for
solutions of the form ($\mu ,\nu = 1,\dots ,4$)
\begin{equation} \label{ansatzg}
ds^2 =
  e^{2A\left(x^5\right)}\bar{g}_{\mu\nu}\left(x^\mu\right)dx^\mu
  dx^\nu + 
  \left(dx^5\right)^2 ,
\end{equation}
and
\begin{equation} \label{ansatzp}
\phi = \phi\left(x^5\right)
\end{equation}
(a non trivial 55 component of the metric which may appear after the
Weyl transformation can be absorbed in a coordinate transformation). 
Further, we chose $\bar{g}_{\mu\nu}$ to be the metric of a maximally
symmetric 3+1 dimensional space
\begin{equation}
\bar{R}_{\mu\nu\kappa\lambda}= \bar{\Lambda}\left(
  \bar{g}_{\mu\kappa} \bar{g}_{\nu\lambda}- \bar{g}_{\mu\lambda}
  \bar{g}_{\nu\kappa} \right) ,
\end{equation}
where the bar indicates that the Riemann tensor is computed with
respect to the metric $\bar{g}_{\mu\nu}$, whose explicit form was given
in Section \ref{section:setup}.

The equations of motion derived from (\ref{brane-frame}) for
$|b|=\frac{4}{3}$ are (as in the previous sections, a prime denotes derivative
with respect to $x^5$)
\begin{equation}
-3\bar{\Lambda}e^{-2A} + 2 A^{\prime\prime} + 5 {A^\prime}^2
 =  0 \label{dilaton}
\end{equation}
\begin{equation} 
-6\bar{\Lambda}e^{-2A}+  6{A^\prime}^2 -
\frac{4b}{|b|}A^\prime\phi^\prime = 0 \label{ff}
\end{equation}
\begin{equation}
-3\bar{\Lambda}e^{-2A}+3A^{\prime\prime} +
6{A^\prime}^2-3\frac{b}{|b|}A^\prime\phi^\prime -
\frac{b}{|b|}\phi^{\prime\prime} + {\phi^\prime}^2
+\frac{1}{2}T
e^{\frac{b}{|b|}\phi}\delta\left(x^5\right)=0,\label{munu}
\end{equation}
where (\ref{dilaton}) is the $\phi$ equation of motion, 
and (\ref{ff}), (\ref{munu}) are the 55, and
$\mu\nu$ components of the Einstein equations, respectively.
We are looking for solutions with continuous functions $A$ and
$\phi$. The advantage of the JBD frame is that the source term appears
only in one of the equations (\ref{munu}). From (\ref{dilaton}) we see
immediately that $A^\prime$ must be continuous such that
$A^{\prime\prime}$ is finite. With (\ref{ff}) follows that either
$\phi^\prime$ is also continuous or $A^\prime (0) = 0$. If $\phi^\prime$
is continuous then (\ref{munu}) implies that $T$ has to vanish - a case
in which we are not interested. Therefore we take $A^\prime (0) = 0$ which 
restricts the value of $\bar{\Lambda}$ to vanish.
So, without solving explicitly any differential equation we rederived
that for $|b|=\frac{4}{3}$ there are no nearby curved solutions. Just
for completeness we give the solution to the remaining equation
(\ref{munu}). Solving in the bulk gives
\begin{equation}\label{solution-flat}
A(x^5) = const\ ,  \qquad
\phi = - \frac{b}{|b|}\log \left| \alpha_i x^5 +\beta_i\right| ,
\end{equation}
where the index $i=1,2$ labels the case of $x^5<0$, $x^5>0$, respectively. 
Using the $Z\!\!\!Z_2$ symmetry of the modulus function we can without
loss of generality restrict to the case that $\alpha_i>0$. Further, we
want to have a finite 4d Planck mass, and this forces us to consider a
solution with a singularity appearing for some value of $x^5$. Then
continuity at zero leads to,
\begin{equation}
\beta_1=-\beta_2=\beta >0.
\end{equation}
The jump condition on the first derivative of $\phi$ at zero finally
restricts 
\begin{equation} \label{jump}
\alpha_1 +\alpha_2 = \frac{1}{2} T
\end{equation}
(for a symmetric solution this implies $\alpha_1 = \alpha_2
= \frac{1}{4}T$).
Transforming back to the Einstein frame and performing an appropriate
coordinate transformation in the fifth direction one finds easily that
this is solution $II$ of \cite{Kachru:2000hf}, or equivalently the
solution discussed in \cite{Arkani-Hamed:2000eg} (i.e. one recovers Eq.
(\ref{eq:Phi_zero}) and (\ref{eq:A_zero}) with
$\epsilon_1 = \epsilon_2 = - \frac{b}{|b|}\, $, $c_1 = - c_2$ and $d_1 = d_2$).

As a next step we want to investigate how stable the constraint
$\bar{\Lambda}=0$ is against deformations of the bulk Lagrangian. To
this end, we will focus on the special case of adding an exponential
bulk potential for the dilaton $\phi$, insisting, however, on
$|b|=\frac{4}{3}$ . (To some extent this mimics
also the option of screening the singularities by additional branes -
the additional bulk potential can be viewed as filling the space with
a continuum of three-branes.) To be specific,
we modify the action (\ref{brane-frame}) as follows,
\begin{equation} \label{lbulkon}
S= \int d^5 x\sqrt{-G}\left[e^{-\frac{3}{4}b\phi} R
    -\Lambda e^{\left(a-\frac{5}{4}b\right)\phi}\right] 
-\int_{x^5=0}d^4x\sqrt{-g} T ,
\end{equation}
which corresponds to adding a bulk potential $V(\phi) = \Lambda e^{a \phi}$
in the original Einstein frame.
The equations of motion with the ansatz (\ref{ansatzg}) ,
(\ref{ansatzp}) read
\begin{equation}
-3\bar{\Lambda}e^{-2A}+2A^{\prime\prime} + 5 {A^\prime}^2
+\frac{5b-4a}{12b} \Lambda e^{\left(a-\frac{b}{2}\right)\phi}
 =  0 \label{dilatonL}
\end{equation}
\begin{equation} 
-6\bar{\Lambda}e^{-2A}+  6{A^\prime}^2 -
\frac{4b}{|b|}A^\prime\phi^\prime +\frac{1}{2}\Lambda
e^{\left(a-\frac{b}{2}\right)\phi} = 0 \label{ffL} 
\end{equation}
\begin{equation}
-3\bar{\Lambda}e^{-2A}+3A^{\prime\prime} +
6{A^\prime}^2-3\frac{b}{|b|}A^\prime\phi^\prime -
\frac{b}{|b|}\phi^{\prime\prime} + {\phi^\prime}^2
+\frac{1}{2}\Lambda
e^{\left(a-\frac{b}{2}\right)\phi}+
\frac{1}{2}Te^{\frac{b}{|b|}\phi}\delta\left(x^5\right)=0.\label{munuL}
\end{equation}
Together with the requirement of having continuous functions $A$ and
$\phi$ the first equation (\ref{dilatonL}) implies that $A^\prime$
must be continuous as well. In order to allow for a non vanishing $T$
in the third equation (\ref{munuL}) we need a jump in $\phi^\prime$ at
$x^5 =0$. From the second equation (\ref{ffL}) one sees that this is
possible only for
\begin{equation}
A^\prime (0) = 0,
\end{equation}
and that any solution has to satisfy the condition
\begin{equation}
6 \bar{\Lambda} e^{-2A(0)} = \frac{1}{2} \Lambda
e^{\left(a-\frac{b}{2}\right)\phi(0)}.
\label{condition}
\end{equation}
This is quite a remarkable result. Insisting on $|b|=\frac{4}{3}$ but
allowing for non-zero $\Lambda$ we find that the effective 4d
cosmological constant $\bar{\Lambda}$ necessarily differs from
zero. The situation is 
somewhat complementary to the case $|b| < \frac{4}{3}$. There nearby
curved solutions exist even for vanishing $\Lambda$
\cite{Kachru:2000xs}. On the other hand it is possible to have flat
solutions for non-vanishing $\Lambda$ (solution $III$ in
\cite{Kachru:2000hf} and generalizations thereof \cite{Csaki:2000wz}).
To complete the discussion, we now solve the equations in the particular
case where the warp factor is constant (which trivially satisfies the
constraint $A'(0)=0$), and show that the corresponding solution gives the
flat solution (\ref{solution-flat}) in the limit $\Lambda
\rightarrow 0$. Plugging $A'(x^5)=0$ into the equations (\ref{dilatonL})
and (\ref{ffL}), one finds that they can be solved only if
\begin{equation}
a = \frac{b}{2} 
\label{choice}
\end{equation}
(conversely, if we choose the bulk scalar potential to satisfy
(\ref{choice}), the equations of motion imply that $A'(x^5)=0$).
Condition (\ref{condition}) then becomes
\begin{equation}
\bar{\Lambda} e^{-2A(0)} = \frac{1}{12}\Lambda.
\end{equation}
For $\Lambda < 0$ we find
\begin{equation} \label{solution-phi}
\phi = \left\{
\begin{array}{l}
-\frac{b}{|b|}\log\left| \frac{\alpha_1\sinh \rho x^5}{\rho}
+\beta\cosh\rho x^5\right|\, x^5 < 0\\
-\frac{b}{|b|}\log\left| \frac{\alpha_2\sinh \rho x^5}{\rho}
-\beta\cosh\rho x^5\right|\, x^5 > 0, \end{array}\right.
\end{equation}
where $\rho = \sqrt{-\frac{\Lambda}{4}}$.
In order to have singularities we chose $\alpha_i, \beta >0$.
The function is continuous, and the jump conditions on the first
derivatives leads to (\ref{jump}). For $\Lambda >0$ we replace in
(\ref{solution-phi}) the hyperbolic functions by their trigonometric
counterparts and $\rho = \sqrt{\frac{\Lambda}{4}}$. The limit
$\Lambda \rightarrow 0$ corresponds to $\rho \rightarrow 0$, and
one easily rediscovers (\ref{solution-flat}).

%%%%%%%%%%%%%%%%%%%%%%%%%%%%%%%%%%%%%%%%%%%%%%%%%%%%%%%%%%%%%%%%%%%%%%%

\section{Further constraints on self-tuning models}

Up to now, we have discussed the self-tuning proposal at the level 
of an effective classical Lagrangian. We did not specify the relation of these 
classical models to a realistic quantum field theory. 
In particular, making such a relation one needs to specify the meaning of 
the brane tension in terms of the perturbative Lagrangian for matter fields,
and any other, nonperturbative in 5d, pieces of the Lagrangian on the brane. 
Firstly, as we argue later on, there may be some primordial nonzero
contribution to the tension on any wall, which is the remnant of the higher 
dimensional theory, and can be considered as nondynamical and nonperturbative 
in five dimensions. Such contributions do not even need to be a part of 
a supersymmetric sigma model on the brane, so they are not protected, or set 
to zero, by brane supersymmetry. Further,  
there is the matter Lagrangian which has perturbative couplings and which is
modified by quantum corrections on the brane. This Lagrangian may be globally
supersymmetric, or may just be  
the one of the usual Standard Model. There are basically two valid  points
 of view on the place of the 
quantum corrections in the effective classical brane tension. One approach  
is based on the observation that there exist (fine-tuned) Poincare 
invariant solutions for a wide range of couplings of the bulk scalar to the 
wall. Then, even if each loop contribution on the brane comes with a different
dependence 
on $\phi$, the expectation value of the sum of them up to any 
arbitrarily choosen order of the perturbation theory can be identified 
with brane tension in our models and efficiently screened. This 
point of view is valid in the present  paper up to the section 4, and we 
take in that part the brane tension to be 
$-e^{b \phi} \int_{M_{(4)}} (<V^{(0)...(n)}_{eff}>+ T_p)$, where $T_p$ is
the primordial brane tension. However, in such a 
situation the Poincare invariant solution is not the only one, there 
exist nearby curved solutions which give nonvanishing cosmological 
constant in 4d. In a more ambitious approach, one would demand, that there 
should be no nearby curved solutions. This, as shown in
\cite{Arkani-Hamed:2000eg}, requires a very careful choice of the the brane
model and its coupling 
to the bulk scalar. There one has to assume that the classical brane tension 
is just the sum of some 
primordial tension and the vacuum value of the classical Lagrangian for fields
living on the brane, and watch explicitly the quantum corrections to see 
whether the required form of the model does not get disturbed.  
To be more specific, we saw in the previous section that if we want to 
achieve certain 
uniqueness of Poincare invariant solutions, it is 
important to have a very particular value of $b$, like $|b|=4/3$ of 
the previous section. It has been argued in \cite{Arkani-Hamed:2000eg} 
that it is possible 
to write down a model where quantum corrections respect the value of $b$ 
once it is set at tree-level. The associated stability  
of the Poincare invariant 
solution with respect to curved deformations is in fact the second part of the 
original self-tuning proposal. Hence it is instructive to take a closer 
look at the 
specific brane model assumed in \cite{Arkani-Hamed:2000eg}. 
The action integral is 
$\int d^5 x \sqrt{-G}( R - \frac{4}{3} (\partial \phi)^2) + 
\int d^5 x \sqrt{-\bar{g}} \delta (x^5) L_b (H; \bar{g}_{\mu \nu})$
where $\bar{g}_{\mu \nu} = g_{\mu \nu} e^{\frac{b}{2} \phi}$, $g_{\mu \nu}$
is the induced metric on the brane and $|b|=4/3$. 
The special feature of this model is the coupling of the scalar field to 
the brane in its equation of motion 
\beq
\frac{8}{3} (\sqrt{-G} \phi')' - \frac{b}{4} \sqrt{-g} \delta (x^5) 
\theta^{\mu}_{\mu} = 0
\eeq
where $\theta^{\mu}_{\mu}$ is the trace of the brane energy momentum tensor,
defined by $\delta \left( \int d^4x \sqrt{-g}\, L_b (H; g_{\mu \nu}) \right)
= - \frac{1}{2} \int d^4x \sqrt{-g}\, \theta_{\mu \nu} \delta g^{\mu \nu}$.
This includes contributions from the dynamical sector 
containing the SM fields, and that from any primordial brane tension $T_p$,
$(\theta^{\mu}_{\mu})_p = 4 T_p e^{b \phi_0}$. Let us note that if we were
in four dimensions, the equation of motion for the scalar which couples
to the trace of the energy-momentum tensor only would automatically imply 
vanishing of that trace on-shell, hence the vanishing of the cosmological 
constant. However, in the present case the vacuum value of derivatives   
of the scalar with respect to the coordinate transverse to the brane 
can be nonzero without violating the Poincare invariance. Hence, the equation 
of motion does not demand the vanishing of the trace of the brane 
energy-momentum tensor. In fact this is the situation we consider in this 
paper. However, the nonvanishing of $\theta^{\mu}_{\mu}$ implies the 
scale anomaly on the brane which in turn implies that the Weyl rescaling 
on the brane is anomalous. This means that the conjectured decoupling 
of the scalar from the brane, hence `conservation' of $b$ by brane physics,
may be violated in the full quantum brane theory. In this respect, we agree
with the argumentation put forward in Ref. \cite{deAlwis:2000qc}.
On the other hand, one could imagine choosing the brane sector which is
conformally invariant to all orders, like $N=4$ Super-Yang-Mills models.
This implies $\theta^{\mu}_{\mu}=0$ at all orders and the background
cosmological constant vanishes as a result of a symmetry. In a realistic
model however, such a symmetry has to be broken, and the cosmological
constant problem reappears. \\

It is interesting to write down explicitly the 1-loop corrections 
to the effective potential in an example of a model conformally coupled
to the scalar.  
To this end we recall that the ultimate 
contribution to the vacuum energy which must be cancelled in order to 
be consistent with observation, arises in low-energy theory, say around the 
$1 \; TeV$ scale. At that scale in all realistic models 4d brane symmetries 
which may have something to do with protecting the vacuum energy, like 
$N=1$ supersymmetry or conformal invariance, are broken. 
This is the place where the mechanism 
of self-tuning (if not fine-tuned) would be really useful. 
To see what happens when quantum effects are taken into account, let us 
specify the representative content of the brane to be given as 
\beq
e^{b \phi} \int \sqrt{-g} (e^{-b/2 \phi} g^{\mu \nu} \partial_\mu \bar{\psi}
\partial_\nu
\psi - m^2 \bar{\psi} \psi + i e^{-b/4 \phi}  e^{\mu}_a \bar{\lambda} \gamma^a
\partial_\mu 
\lambda -M \bar{\lambda} \lambda -\frac{1}{4 g^2} e^{- b \phi} F^2 -V_0-T_p ).
\eeq 
To compute one-loop contribution to the effective potential we first perform 
the redefinition of the fields so that their kinetic terms are canonical 
with respect to Minkowski background on the brane, and then use the standard 
formulae with a physical UV cut-off $\Lambda$ which we understand to be close
to $1 \; TeV$. After going back to 
the original frame  one obtains $-\int \sqrt{-g} ( e^{b \phi} (V_0+ T_p)
+ Str {\bf 1}\, {\cal O} (\Lambda^4)  
+ \frac{1}{32 \pi^2}
 e^{b/2 \phi} \Lambda^2 (m^2 - M^2) ) + ... $, where 
one can see terms which are multiplied by different powers of $e^{b \phi}$.
Of course, this situation is also manifest in models without conformal 
coupling of the brane to the scalar field. 
Hence, in general, even if one carefully adjusts $|b|=4/3$, at tree level, 
one departs fom this choice after including quantum corrections. 
This simply means that the classical 
selection rules have limited value in the true, quantum, world.

We want to mention another troublesome feature of the model with conformal 
coupling of the 
scalar to the branes, which is an inconsistency with long range properties 
of 4d 
gravity at the classical level. This becomes clear in the 
Jordan-Brans-Dicke frame introduced in the section 4. 
In that frame the model of 
\cite{Arkani-Hamed:2000eg} takes the form of a generic 
Jordan-Brans-Dicke model,
and is subject to constraints given by the analysis of 4d Jordan-Brans-Dicke 
systems performed 
in \cite{Damour:1993id}. The 4d 
Jordan-Brans-Dicke action in the JBD frame looks as follows:
$\int d^4 x \sqrt{- g} ( \frac{\Phi^2}{2} R - 2 \omega (\Phi) g^{\mu \nu}
\partial_\mu 
\Phi \partial_\nu \Phi - V(\Phi) )$ with the assumption that matter fields 
are coupled only to the metric $g_{\mu \nu}$ and not to $\Phi$. This is very 
similar to the Lagrangian given in (\ref{brane-frame} ), up to a redefinition
of the field 
$\phi$ and integration over the fifth dimension. 
It is easy to see,  
through the integration of the fifth dimension directly in the 5d JBD frame, 
that in the case of (\ref{brane-frame} ) $\omega = 0$. 
It has been shown in \cite{Damour:1993id} that there exist 
constraints on the 
value of the parameter $\omega$ 
of JBD models. This constraint says that 
$\frac{1}{2 \omega + 3} \; < \; 10^{-3}$, 
hence in the model which we are discussing here the above 
bound is violated. The troublesome constraint could be avoided if the 4d JBD
scalar would receive a mass of the order of 
$1/mm$. This can be achieved by introducing a very soft potential 
for the scalar field, so that the experimental bound on the 4d 
cosmological constant, $\rho_\Lambda < (0.002\; eV)^4$ is not violated. 
At this point however, we should recall the result of the preceding section,
where we have shown that introducing a simple bulk potential while insisting
on $|b|=4/3$ has led one from a flat to a curved solution. \\

However, one should repeat, that
in the context of generic models lacking  conformal 
coupling of branes to a bulk scalar, it is still likely 
that even for a modified, so pretty generic, dependence on $\phi$ there 
exists a Poincare invariant vacuum background. 
The problematic point is, that in response to quantum corrections 
the tensions on the additional branes located 
at singularities must get re-finetuned. This would have to be achieved   
by some mechanism operating at low 
energies in addition to the bulk scalar mediation. 
One may speculate that the bulk KK modes   
could be an agent of such an readjustment, but a separate calculation 
is needed to decide whether this is the case. \\

The next issue which needs to be addressed is the physics at or near branes
located at singularities. One way to conceal this problem is to say that the 
singularities are resolved in the microscopic higher dimensional, $D>5$,
model. 
However, perhaps one can still make some useful observations looking more 
carefully at the 5d models. A way to test physics near singularities 
is to perturb the model by a small point-like test  mass and to read off the
strength of interaction of this test mass  with the 
gravitational field. 
The action for such a test 
particle  is $S_p = -m \int ds $ in 5d coordinates and it gives a contribution 
to the energy momentum tensor of the form $T^{MN} = \frac{m}{\sqrt{-g}}
\int d \tau \frac{ {\dot{x}}^M {\dot{x}}^N}{\sqrt{ - g_{PQ} 
{\dot{x}}^P {\dot{x}}^Q }} \delta (x - x(\tau)) $. The perturbation of the
action  under a small variation $\delta h_{\mu \nu}$ is 
$\delta S_p = - \int d^5 x \sqrt{-g} (e^{2 A} \delta h_{\mu \nu})=
- \int d^4 x \sqrt{-\tilde{g}} \delta h_{\mu \nu} \tilde{T}^{\mu \nu}$ where 
$\tilde{T}^{\mu \nu} = \int dy e^{6 A} T^{\mu \nu}$. One obtains then 
the 4d energy momentum tensor due to a test mass $\tilde{T}^{\mu \nu} =
\frac{m e^{A(x^5)}}{\sqrt{-\tilde{g}}}
\int d \tau \frac{ {\dot{x}}^{\mu} {\dot{x}}^{\nu}}{\sqrt{ - \tilde{g}_{PQ} 
{\dot{x}}^P {\dot{x}}^Q }} \delta (x - x(\tau)) $. 
This means that the mass which interacts with the gravitational field 
at the location $x^5$ is $m e^{A(x^5)}$, hence, for instance, it vanishes 
whenever $e^A =0$. Hence, physical matter located at singular branes 
is effectively massless. Essentially, since $\sqrt{-G}=0 $ at the singular
brane, local dynamics there is also suppressed. Then the  mechanism 
which creates correlations between vacuum tensions on physical and singular 
branes might be the result of some higher dimensional consistency condition, 
rather than a dynamical one. For instance, one can recall 
the way the Bianchi identity is fullfilled in the Horava-Witten model. 
There one needs to switch on vacuum fluxes  
of the gauge fields along the internal dimensions. The fluxes are different 
on different walls, but 
their sum is fixed by the Bianchi identity. These fluxes, together with 
that of gravitational curvature tensor in six compact dimensions, lead 
to nonzero but correlated brane tensions in five space-time dimensions.
  
To learn something about dynamics in singular warped universe,
it is useful to compute in the nonrelativistic limit 
the force acting on a test mass $m$. 
From the geodesic equation one easily obtains the acceleration of the 
freely  falling particle $\frac{d^2 X^M}{d t^2} = - \Gamma^{M}_{00} 
= - \frac{1}{2} g^{MN} \frac{\partial g_{00}}{\partial x^N}$. 
This is nonvanishing along the fifth direction and gives 
$\frac{d^2 X^5}{d t^2} = A'(x^5) e^{2 A(x^5)}$. Hence, the force acting on 
the slowly moving test particle is $F^5 = m  A'(x^5) e^{2 A(x^5)}$.
To see the possible implications let us take an example of the singular 
background which has been  discussed in the third section of this  paper, 
in the vicinity of the 
naked singularity on the positive side of the $x^5$ axis,
$e^A = e^{1/3 d} | 4/3 y -c|^{1/4}$. When the particle approaches the 
singularity from the left, it feels the force 
$F^5 \approx - | 4/3 y -c |^{-1/2}$ which repells the test particle away 
from the singularity. This may be interpreted as a hint at sort of stability
of the system, in the sense that one does not expect the flow of 
physical matter from  the positive-tension brane to the singular negative
tension branes. 

Finally, let us comment briefly on the 
stabilization of the fifth dimension in the 
static (flat) case. The standard way to formulate this problem is to 
separate a $x^5$-independent part of $G_{55}$, let us call it here 
$r_c (x)$, and to require that after 
integrating out all remaining degrees of freedom and integrating over 
$x^5$ there remains an effective potential for $r_c$ such that either 
it has a minimum (and thus truly stabilizes $r_c$) or it is flat, 
so no effective force moves $r_c$ around. This second possibility 
would correspond for instance to a modulus in a supersymmetric model, but 
we know that at low energies supersymmetry needs to be completly broken down. 
Hence, of actual interest is the first one. The examples of that favourable 
situation are the models (with repaired singularities) presented in this paper
which contain a nontrivial bulk potential for the scalar field. 
In the case of the models with no potential in the bulk, the distance 
to the singularities was not fixed. Moreover, if one puts the brane screening
the singularity between the SM brane and singularity, its position is also
undetermined. According to what has been said earlier, we expect 
quantum corrections to modify the classical solution in these cases.   
Parenthetically, we would like to notice 
that in the 5d Einstein frame in which we work in the initial sections of 
the paper, the substitute of the equation of motion 
determining $r_c$ is simply 
the Einstein equation $E_{55} - \frac{1}{2} T_{55}= 0$.
In addition, for the general class of solutions discussed here, it has been
shown in Ref. \cite{DeWolfe} that they are stable in the sense that there are
no tachyonic fluctuations.

The remaining constraint on singular warped universes which we want to
mention arises when one computes the corrections to Newton's law due
to the exchange of heavy gravitons \cite{Matyszkiewicz}. 
When one tries to send the singularities far apart, i.e. 
when one considers a macroscopic fifth dimension, then the correction 
to the effective 4d Newton's law is $\delta V = - \frac{1}{r^2} \mu G_N 
\frac{|c_1| + |c_2|}{2 \pi} $. Hence, the correction grows linearly 
with the size of the fifth dimension, similarly to what would happen in 
a toroidal compactification. This means that the distance between 
singularities should be smaller than, say, $1 \; mm$. In this regime 
the correction due to heavy gravitons is exponentially suppressed as
$\delta V = -\frac{1}{2 r} \mu G_N e^{- \frac{\pi r}{2 (|c_1| + |c_2|)}}
(1 + \sin (\frac{ \pi |c_1|}{|c_1| + |c_2|} ))$.

%%%%%%%%%%%%%%%%%%%%%%%%%%%%%%%%%%%%%%%%%%%%%%%%%%%%%%%%%%%%%%%%%%%%%%

\section{Summary and Conclusions}

In the present paper we have argued that in brane worlds the
cosmological constant problem is merely seen from a new
perspective. The degree of fine-tuning needed to obtain a value in
agreement with observational bounds is not improved. We gave explicit
consistency conditions on warped compactifications. These consistency
conditions were derived from the requirement of having a globally well
defined solution to the equations of motion. If the conditions are
violated the metric on the brane is not at a stationary point of 
the 4d effective action. Various examples with singularities are
inconsistent unless the singularities are screened by additional
sources. The vacuum energy at these sources needs to be fine-tuned.

Another issue appearing in the literature is that for $\left| b\right| =4/3$ 
and vanishing bulk potential only
solutions with zero cosmological constant
exist \cite{Arkani-Hamed:2000eg,Kachru:2000xs}.  
We pointed out that
this advantage turns into a disadvantage as soon as a bulk potential
for the scalar is switched on. Then only solutions with a non-zero
cosmological constant exist (the numerical value is given by
parameters of the bulk potential). Hence, in addition of fine-tuning
$\left| b\right| =4/3$ one needs to fine-tune parameters in the bulk
potential even at the classical level.  

Finally, we pointed out that quantum corrections due to the theory
living on a brane may alter the functional form of the coupling of the
scalar to the brane. Other open problems are the stabilization of the
scalar and the inter-brane distance.

%%%%%%%%%%%%%%%%%%%%%%%%%%%%%%%%%%%%%%%%%%%%%%%%%%%%%%%%%%%%%%%%%%%%%%%

\vskip 1cm  
  
\noindent {\bf Acknowledgments}  
  
\noindent We thank Christophe Grojean for comments.
This work has been supported by TMR programs
ERBFMRX--CT96--0045 and CT96--0090.
Z.L. is additionaly supported
by the Polish Committee for Scientific Research grant 
2 P03B 05216(99-2000).  
S.F.\ acknowledges the hospitality of the ESI for Mathematical Physics
in Vienna.

%%%%%%%%%%%%%%%%%%%%%%%%%%%%%%%%%%%%%%%%%%%%%%%%%%%%%%%%%%%%%%%%%%%%%%%

\end{document}